\documentclass[oldversion, printer]{aa}

\usepackage{graphicx}
\usepackage{txfonts}
\usepackage{natbib}
\bibpunct{(}{)}{;}{a}{}{,}

\newcommand{\ii}{\mathrm{i}}                                   
\newcommand{\dd}{\mathrm{d}}                                   
\newcommand{\ee}{\mathrm{e}}                                   
\newcommand{\cc}[1]{{\bar{#1}}}                                
\newcommand{\multind}[1]{{\vec{#1}}}                           
\newcommand{\der}[2]{\frac{\,\dd #1}{\,\dd #2}}                
\newcommand{\rad}{\mathrm{r}}
\renewcommand{\vert}{\mathrm{v}}

\newcommand{\ie}{i.e.\ }
\newcommand{\eg}{e.g.\ }
\newcommand{\citeq}[1]{(\ref{#1})}                             

\begin{document}

\title{Weak nonlinear coupling between epicyclic modes in slender tori}

\author{Ji\v{r}\'{\i}~Hor\'ak}

\offprints{horak@astro.cas.cz}

\institute{Astronomical Institute v.v.i., Academy of Sciences, 
   Bo\v{c}n\'{\i}~II, CZ-141\,31~Prague, Czech Republic}

\authorrunning{J.~Hor\'ak}

\titlerunning{Coupling of epicyclic modes in slender tori}

\date{Received .. ............. 2007; accepted .. ............. 2007}

\abstract{
We examine nonlinear oscillations of slender tori in the vicinity of black holes and compact stars. These tori represent useful probes of the complicated, nonlinear dynamics of real accretion disks and provide at least qualitative understanding of their oscillations. We demonstrate that epicyclic modes of such tori are weakly coupled due to the pressure and gravitational forces. We explore all possible resonances between two epicyclic modes up to the fourth order. We show that the strongest resonance between axisymmetric modes is 3:2. In addition, any resonance between an axisymmetric and a non-axisymmetric mode is excluded due to axial and equatorial-plane symmetries of the equilibrium torus. We examine a parametric excitation of vertical axisymmetric oscillations by radial oscillations in the 3:2 resonance. We show that the resonance may be significant only for high-amplitude radial oscillations.}


\maketitle


\section{Introduction}   

In the past thirty years, the oscillations of fluid tori orbiting around massive compact objects were studied systematically, particularly in the context of the stability of thick accretion disks \citep[][and many others]{Papaloizou+Pringle1984, Papaloizou+Pringle1985, Blaes1985, Goldreich+1986, Goodman+1987, Narayan+1987}. Torus oscillations have attracted renewed attention following the discovery of kilohertz double-peak quasi-periodic oscillations (QPOs) in the light curves of several accreting black holes (see \citealp{McClintock+Remillard2006} for a review). Frequencies of these oscillations are often in the 3:2 ratio and they are detected mainly when a source is in a so-called steep power-law state. The geometry of the accretion flow in this state is not yet clear, nevertheless a fluid torus created by substantial pressure gradients in the inner parts of the accretion disk and supporting discrete trapped oscillations is a plausible possibility. Such tori appear in many non-radiative global simulations as natural components of magnetohydrodynamic turbulent accretion flows \citep{Hawley+Balbus2002, DeVilliers+2003, Machida+2006}. 

In some models, QPOs are identified with two different modes of torus oscillations. For example, \citet{Rezzolla+2003} described QPOs as the two lowest-order $p$-mode oscillations of a polytropic torus of small thickness with constant angular-momentum distribution. Similarly, \citet{Blaes+2006} explored linear oscillations in slender tori and identified the twin-peak QPOs with the vertical epicyclic mode and the breathing mode. Although frequencies of these modes depend on the position of the torus in the accretion flow, their ratio remains close to 3:2 in a wide range of radii.

In this context, \citet{Abramowicz+Kluzniak2001} proposed an interesting general idea that a nonlinear resonance between two modes of accretion-disk oscillations is responsible for observed QPOs in both black-hole and neutron-star sources. Being more specific, they identified them with the radial and vertical epicyclic modes of accretion tori \citep{Kluzniak+Abramowicz2002} and, adopting the approximation of a slender torus, they showed that the existence of these modes is quite  independent of the torus equation of state \citep{Abramowicz+2006}. Later, \citet{Blaes+2007} demonstrated that the epicyclic modes also exist in thicker polytropic tori and found approximate expressions for their eigenfunctions and eigenfrequencies. 

The resonance arises due to nonlinear coupling between the epicyclic modes. In the scenario suggested by \citet{Kluzniak+Abramowicz2002}, one mode is first excited externally, either by an external periodic forcing due to rotation of the central object \citep{Lee+2004, Lee2005} or by a stochastic forcing due to turbulence in the accretion flow \citep{Vio+2006}. The other mode is then continously supplied by the energy from the former one via the parametric resonance. The nonlinear interaction may be strong enough and the amplitude of the second mode may eventually exceed the amplitude of the first mode. In the strong gravitational field, the strongest resonance occur when the frequency ratio of the vertical and radial oscillations is close to 3:2.

In the context of QPOs, the resonance effects in thin accretion disk were already studied by \citet[][and references therein]{Kato2003, Kato2004, Kato2008}. In these models, QPOs are identified with waves (corresponding to either $g$-mode or $p$-mode oscillations), parametrically excited by the deformation of the disk (warp or precession). Mutual nonlinear interactions between different modes of a thin disk were recently studied also by \citet{Fogelstrom+2008} using a local approximation.

In this note, we further examine the importance of nonlinear coupling between epicyclic oscillations of the slender torus. Since the geodesic equations are separable, earlier studies, based on epicyclic motion of test particles, used an additional `ad-hoc' force to initiate the resonance \citep{Abramowicz+2003, Rebusco2004}. We show that in the case of epicyclic modes of fluid torus, the situation differs and both epicyclic oscillations are naturally coupled due to pressure and gravity. We also estimate the strength of this coupling. 

The plan of the paper is following. In Sect.~\ref{sec:form}, we briefly introduce the approximation of slender tori and review the formalism that we use to calculate nonlinear interactions among slender torus modes. This formalism is similar to that used frequently for modeling nonlinear oscillations of rotating stars \citep[e.g.][]{Dyson+Schutz1979, Schutz1980a, Schutz1980b, Kumar+Goldreich1989, Wu+Goldreich2001, Schenk+2002, Arras+2003}. Section~\ref{sec:cplcoeff} describes the coupling coefficients. We estimate the relative importance of pressure and gravitational coupling in different types of torus oscillations. In Sect.~\ref{sec:resonances} we discuss possible resonances between epicyclic modes based on their symmetry properties, and describe a particular example of the 3:2 epicyclic resonance. Sections \ref{sec:discusion} and \ref{sec:conclusions} are devoted to discussion and our conclusions.

\section{Formalism} \label{sec:form}

A stationary configuration of the Newtonian slender torus in a general axisymmetric gravitational field $\Phi(r,z)$ is described by \citet{Blaes+2007}. The torus consists of a polytropic fluid with the equation of state $p\propto\rho^{1+1/n}$, where $p$ and $\rho$ are the pressure and the mass-density at a given point, and $n$ is the polytropic index. In cylindrical coordinates $\{r,\phi,z\}$, the velocity of the stationary unperturbed flow is purely azimuthal $\vec{v}=\Omega\vec{e}_{\phi}$ ($\vec{e}_\phi$ is the unit vector in the azimuthal direction) and constant at cylinders $r=\mathrm{const}$, i.e.\ $\Omega=\Omega(r)$. Furthermore, the torus is symmetric with respect to the equatorial plane. The density and pressure profiles of the torus can be expressed as  $\rho(\vec{x})=\rho_0 f^n(\vec{x})$ and $p(\vec{x})=p_0 f^{n+1}(\vec{x})$, where $\rho_0$ and $p_0$ are the density and pressure at the center of the torus (the circle $r=r_0$; hereafter the subscript `0' refers to an evaluation at this point) and $f(\vec{x})$ is an auxiliary function ($f=0$ and $1$ correspond to the boundary and the center of the torus). The size of the torus depends on the slenderness parameter $\beta$, which is defined by $\beta^2= 2(n+1)p_0/(\rho_0 r_0^2\Omega_0^2)$. The azimuthal velocity at the torus center is given by the local Keplerian angular frequency $\Omega_0=(r^{-1}\partial_r\Phi)_0$. 

In the limit of $\beta\rightarrow0$, the torus with a constant-angular-momentum distribution has an elliptical cross-section that have lengths of the major and minor axes in the ratio of the radial and vertical epicyclic frequencies measured at the center of the torus, $\omega_r^2 = (\partial^2_r\Phi)_0 + 3\Omega_0^2$ and $\omega_z^2=(\partial_z^2\Phi)_0$. The distance from the outer or inner edge to the center of the torus is $\Delta r=\beta r_0/\bar{\omega}_r + \mathcal{O}(\beta^2)$, where $\bar{\omega}_r \equiv\omega_{r}/\Omega_0$. 

\subsection{Linear perturbation} 
\label{subsec:lin}

Since the equilibrium configuration is stationary and axisymmetric, all linear  perturbations are proportional to $\exp[-\ii(\omega t - m\phi)]$, where $\omega\in\mathbb{C}$ and $m\in\mathbb{Z}$ are the eigenfrequency and the azimuthal wavenumber of the perturbation. The eigenfunctions of the torus are traditionally expressed in terms of the \citet{Papaloizou+Pringle1984} variable $W\equiv-\delta p/(\sigma\rho)$, where $\delta p$ is the Eulerian pressure perturbation and $\sigma\equiv\omega-m\Omega$ is the eigenfrequency measured in the system comoving with the equilibrium flow. In next, these functions are referred to as the Eulerian eigenfunctions.

Alternatively, the same perturbations can be expressed in terms of the Lagrangian displacement
\begin{equation}
  \vec{\xi} = \left\{ 
    \frac{m\hat{\kappa}^2 W - 2\sigma\Omega\,r\,\partial_r W}{
      2\Omega r(\sigma^2-\hat{\kappa}^2)},
    -\frac{\ii(m\sigma W - 2\Omega\,r\,\partial_r W)}{
      (\sigma^2-\hat{\kappa}^2)r},
    -\frac{\partial_z W}{\sigma}
    \right\},
    \label{eq:eula}
\end{equation}
where $\hat{\kappa}^2=r_0^{-3}\dd\ell^2/\dd r$ and $\ell(r)$ is the angular-momentum distribution, $\ell(r)=r^2\Omega(r)$. The vector field $\vec{\xi}$ is referred to as the Lagrangian eigenfunction\footnote{The differences between equation (\ref{eq:eula}) and the equation (2.14) of \citet{Papaloizou+Pringle1985} are due to a different definition of the corotation frequency $\sigma$}. 

For infinitely slender torus, the Eulerian eigenfunctions of the radial and vertical epicyclic modes are $W_\rad\propto(r-r_0)\exp(\ii m\phi)$ and $W_\vert\propto z\exp(\ii m\phi)$ \citep[see][]{Blaes+2006}. Each of the eigenfunctions corresponds to both a positive and negative corotation eigenfrequency of the mode, $\sigma_0 = \pm\omega_r$ and $\sigma_0=\pm\omega_z$, respectively. In the Lagrangian description, the eigenfunctions and eigenfrequencies of the epicyclic modes are given by 
\begin{equation}
  \vec{\xi}_\rad = \mathcal{A}_\rad
    \left\{1,\,\mp\frac{2\ii\Omega_0}{\omega_r r_0},\,0
    \right\}\ee^{\ii m\phi}, 
  \quad
  \omega_\rad = m\Omega_0\pm\omega_r
  \label{eq:rad}
\end{equation}
and
\begin{equation}
  \vec{\xi}_\vert = \mathcal{A}_\vert
    \left\{0,\,0,\,1\right\}\ee^{\ii m\phi},
  \quad
  \omega_\vert = m\Omega_0\pm\omega_z\,,  
  \label{eq:vert}
\end{equation}
where $\mathcal{A}_\rad$ and $\mathcal{A}_\vert$ are normalization real constants. The eigenfunctions of the two modes with opposite eigenfrequencies are complex conjugated, \ie if $(\omega,\vec{\xi})$ represents the solution to the eigenvalue problem then another solution is $(-\omega,\vec{\xi}^\ast)$.

\subsection{Nonlinear coupling} 

The nonlinear oscillations of polytropic torus are governed by the partial differential equation
\begin{equation}
  \frac{\mathrm{D}^2\xi^i}{\mathrm{D}t^2} - \frac{1}{\rho}
  \nabla_j\left[(\gamma-1)p(\nabla\cdot\vec{\xi})g^{ij} + 
  p\nabla^i\xi^j\right] + \xi^j\nabla_j\nabla^i\Phi = a_\mathrm{n}^i(\vec{\xi}),
  \label{eq:gov_xi}
\end{equation}
where $\mathrm{D}/\mathrm{D}t=\partial_t + v^k\nabla_k$ is the Lagrangian flow derivative, $\gamma=1+1/n$ is another polytropic index and $g^{ij}$ denotes contravariant components of the metric tensor. All nonlinear terms are absorbed into the acceleration term $a_\mathrm{n}^i(\vec{\xi})$ on the right-hand side of the equation.

At a given time, the Lagrangian displacement of the nonlinear oscillations can be expressed as a linear combination of the Lagrangian eigenfunctions $\vec{\xi}_A$\footnote{In the case of Jordan chain modes the expansion \citeq{eq:xi} must be completed by the associated Jordan-chain vectors \citep[see][]{Schenk+2002}.},  
\begin{equation}
  \vec{\xi}(t,\vec{x}) = \sum_A c_A^\ast(t)\,\vec{\xi}_A^\ast(\vec{x}), 
  \label{eq:xi} 
\end{equation}
where the capital Latin indices symbolize different modes. The coefficients $c_A$ describe the time-dependent  instantaneous amplitudes and phases of different linear modes. When this form is substituted into Eq.~\citeq{eq:gov_xi}, we obtain a system of many oscillators 
\begin{equation}
  \der{c_A}{t} + \ii\omega_A c_A=  
  \frac{\ii}{b_A} \mathcal{F}^\ast_A(c_I).
  \label{eq:gov}
\end{equation}
Different oscillators are coupled by nonlinear functions $\mathcal{F}_A(c_I)$. In the absence of nonlinearities, the solutions of Eq.~(\ref{eq:gov}) are harmonic functions $c_A\sim\exp(\omega_A t)$ and Eq.~(\ref{eq:xi}) becomes just a superposition of linear modes. The coefficients $b_A$ depend solely on $\vec{\xi}_A$ and takes into account the normalizations of the modal eigenfunctions. They are given by \citep{Schutz1980a, Schenk+2002},
\begin{equation}
  b_A = 2\int_V\rho
  \left\{\sigma_A\left[(\xi_A^r)^2 + (\xi_A^z)^2 + (r\xi_A^\phi)^2\right]
  - 2\ii r\Omega\xi_A^r\xi_A^\phi\right\}\dd V,
  \label{eq:b}
\end{equation}
where $V$ is the torus volume. The nonlinear coupling functions can be expanded into a Taylor series,
\begin{eqnarray}
  \mathcal{F}_A &=& \sum_{B,C}\kappa_{ABC}\,c_B c_C + \sum_{B,C,D}\kappa_{ABCD}\,c_B c_C c_D + 
  \nonumber\\ &\phantom{=}& 
  \sum_{B,C,D,E}\kappa_{ABCDE}\,c_B c_C c_D c_E + \dots\,.
  \label{eq:nonlin}
\end{eqnarray}
The coefficients of this expansion $\kappa_{ABC}$, $\kappa_{ABCD}$ and $\kappa_{ABCDE}$ are the three-mode, four-mode and five-mode coupling coefficients. These are studied in detail in the following section.

Equation \citeq{eq:gov} is not applicable to Jordan-chain modes, for which $b_A=0$. This is not however the case for the epicyclic modes. It can be verified using Eqs.~\citeq{eq:rad}, \citeq{eq:vert} and \citeq{eq:b} that
\begin{equation}
  b_\rad = \pm 2\omega_r\mathcal{M}\mathcal{A}_\rad^2,
  \quad
  b_\vert = \pm 2\omega_z\mathcal{M}\mathcal{A}_\vert^2,
\end{equation}
where
\begin{equation}
  \mathcal{M}=\int_V\rho(\vec{x})\dd V =
  \frac{2\pi^2\beta^2 r_0^3\rho_0}{(n+1)\,\bar{\omega}_r\bar{\omega}_z}
\end{equation}
is the torus mass. In the limit of an infinitely slender torus, $b_A=0$ for corotation modes. They correspond to Jordan chains of length 1 and become unstable when $\beta>0$, developing the Papaloizou-Pringle instability \citep{Blaes1985}. This agrees with the general theory of \citet{Schutz1980b}. 

The decomposition of the nonlinear solutions into eigenfunctions of linear modes is a common procedure in the theory of stellar oscillations. Up to the second order in coefficients $c_A$, our Eqs.~\citeq{eq:xi}, \citeq{eq:gov} and \citeq{eq:nonlin} are identical to Eqs.~(4.12) and (4.13) of \citet{Schenk+2002}, who considered only three-mode nonlinear interactions. 

In principle, a similar approach can be applied to a more general case of MHD flow. An equation that models the nonlinear evolution of the oscillation modes, would have a similar form as Eq.~\citeq{eq:gov}; the specific expression for the coupling coefficients would, however, differ from the present case.

\section{Coupling coefficients} 
\label{sec:cplcoeff} 

Individual fluid elements of the torus move under the combined influence of the gravitational and pressure forces. For this reason, it is beneficial to separate the contributions of pressure and gravity and express coupling coefficients, in general, as
\begin{equation}
  \kappa\equiv\kappa^{(\mathrm{p})} + \kappa^{(\mathrm{g})}\,,
\end{equation}
These two contributions are referred to as the pressure-coupling and gravitational-coupling coefficients. Up to the fourth order, they are given by 
\begin{eqnarray}
  \kappa_{ABC}^{(\mathrm{p})}\!\!&=&\!\!\frac{1}{2}\int_V p\,\Big\{
  (\gamma-1)^2\eta_A\eta_B\eta_C + 3 (\gamma-1)\eta_{(A}\eta_{BC)}+
  \nonumber\\ \!\!&\phantom{=}&\!\!
  2\eta_{(ABC)}\Big\}\,\dd V,
  \label{eq:kappa3p}
  \\
  \kappa_{ABC}^{(\mathrm{g})}\!\!&=&\!\!-\frac{1}{2}\int_V 
  \rho\,\xi^i_A\xi^j_B\xi^k_C\nabla_i\nabla_j\nabla_k\Phi\,\dd V,
  \label{eq:kappa3g}
\end{eqnarray}
\begin{eqnarray}
  \kappa_{ABCD}^{(\mathrm{p})}\!\!&=&\!\!
  -\frac{1}{3!}\int_V \gamma p\,\Big\{
  (3-3\gamma+\gamma^2)\,\eta_A\eta_B\eta_C\eta_D + 
  8\,\eta_{(A}\eta_{BCD)} +
  \nonumber\\ \!\!&\phantom{=}&\!\!
  6(\gamma-2)\,\eta_{(A}\eta_B\eta_{CD)}+
  3\eta_{(AB}\eta_{CD)}\Big\}\,\dd V
  \label{eq:kappa4p}
  \\
  \kappa_{ABCD}^{(\mathrm{g})}\!\!&=&\!\!-\frac{1}{3!}\int_V 
  \rho\,\xi^i_A\xi^j_B\xi^k_C\xi^l_D \nabla_i\nabla_j\nabla_k
  \nabla_l\Phi \,\dd V
  \label{eq:kappa4g}
\end{eqnarray}
and
\begin{eqnarray}
  \kappa_{ABCDE}^{(\mathrm{p})}\!\!&=&\!\!
  \frac{1}{4!}\int_V \gamma p\,\Big\{
  (1+6\gamma-4\gamma^2+3\gamma^3) 
  \eta_A\eta_B\eta_C\eta_D\eta_E +
  \nonumber\\ \!\!&\phantom{=}&\!\!
  10\gamma(\gamma-3) \eta_{(A}\eta_B\eta_C\eta_{DE)} +
  20\gamma \eta_{(A}\eta_B\eta_{CDE)} +
  \nonumber\\ \!\!&\phantom{=}&\!\!
  15(\gamma-1) \eta_{(A}\eta_{BC}\eta_{DE)} +
  20 \eta_{(AB}\eta_{CDE)}\Big\}\,\dd V,
  \label{eq:kappa5p}
  \\
  \kappa_{ABCDE}^{(\mathrm{g})}\!\!&=&\!\!-\frac{1}{4!}\int_V 
  \rho\,\xi^i_A\xi^k_B\xi^l_C\xi^m_D\xi^n_E
  \nabla_i\nabla_k\nabla_l\nabla_m\nabla_n\Phi\,\dd V,
  \label{eq:kappa5g}
\end{eqnarray}
where brackets in the indices denote the symmetrization. For a simpler notation, we introduce scalars
\begin{eqnarray}
  \eta_A &\equiv& 
    (\nabla_i\xi_A^i), \\
  \eta_{AB} &\equiv& 
    (\nabla_i\xi^j_A)(\nabla_j\xi_B^i) \\
  \eta_{ABC} &\equiv& 
    (\nabla_i\xi^j_A)(\nabla_j\xi_B^k)(\nabla_k\xi_C^i).
\end{eqnarray} 

The three-mode pressure-coupling coefficients were derived by \citet{Dziembowski1982}, \citet{Kumar+Goldreich1989}, and reviewed recently by \citet{Schenk+2002}. The four-mode coupling coefficients were derived by \citet{VanHolst+Smeyers1993} \citep[see also][]{VanHolst1994} who considered the more general case of the self-gravitating isothermal star. 

Before calculating exact values, we review the necessary conditions for coupling coefficients to be nonzero and explore the importance of different kinds of coupling for different types of torus oscillations. For this purpose, it is useful to introduce a multi-index notation; the multi-indices are denoted by bold-face letters,  $X_{\multind{A}}=X_{A_1\dots A_n}$ and their absolute values are given by the number of the indices, $|\multind{A}|=n$.

\subsection{Selection rules}

The necessary conditions for non-zero coupling coefficients follow from the symmetry properties of the integrands in Eqs.~\citeq{eq:kappa3p}--\citeq{eq:kappa5g}, and are natural generalization of the well-known selection rules for three-mode coupling \citep[see \eg][]{Schenk+2002}. The \emph{azimuthal selection rule} states that the integrands cannot depend on the azimuthal angle $\phi$,
\begin{equation}
  \sum_{k=1}^{|\multind{A}|} m_{A_k} \neq 0
  \,\,\Rightarrow\,\,
  \kappa_\multind{A}=0.
  \label{eq:rulephi}
\end{equation}
Similarly, as the equilibrium configuration is symmetric with respect to the equatorial plane, the integrations are performed over symmetric intervals in the vertical direction and the coupling coefficients vanish when the integrands are odd functions of $z$. This happens when an odd number of modes is involved, whose Lagrangian eigenfunctions are antisymmetric with respect to the reflection $z\leftrightarrow -z$,
\begin{equation}
  \prod_{k=1}^{|\multind{A}|} \epsilon_{A_k} = -1
  \,\,\Rightarrow\,\,
  \kappa_\multind{A}=0.
  \label{eq:rulez}
\end{equation}
In the above equation, $\epsilon_{A_k}$ is the parity of the $A_k$-mode under reflection about the equatorial plane, and $\epsilon_{A_k}=\pm1$ corresponds to even and odd modes, respectively. The condition \citeq{eq:rulez} is referred to as the \emph{vertical selection rule}.

\subsection{Nodal modes}

We consider modes whose Lagrangian eigenfunctions have at least one (but still less than $1/\beta$)  node in both the $r$ and $z$ directions; such modes are referred to as the \emph{nodal modes}. The Lagrangian displacements and their gradients satisfy the scalings
\begin{equation}
  \vec{\xi}_A \sim \mathcal{A}_A,
  \quad
  \vec{\nabla}\vec{\xi}_A \sim \mathcal{A}_A/(\beta r_0).
\end{equation}
The coupling coefficients can be estimated by
\begin{equation}
  \kappa^{(\mathrm{p})}_\multind{A}\sim p_0 V\,
  [\mathcal{A}/(\beta r_0)]^{|\multind{A}|},
  \quad
  \kappa^{(\mathrm{g})}_\multind{A}\sim \rho_0 V\, 
  \Phi_0\,[\mathcal{A}/r_0]^{|\multind{A}|},
\end{equation}
where the torus volume and the central pressure are
\begin{equation}
  V\sim\beta^2 r_0^3,
  \quad
  p_0\sim\beta^2 r_0^2\Omega_0^2\,\rho_0
\end{equation}
Therefore, the gravitational coupling coefficients are smaller than the pressure ones by a factor proportional to
\begin{equation}
  \kappa^{(\mathrm{g})}_\multind{A}/\kappa^{(\mathrm{p})}_\multind{A}\sim
  \beta^{|\multind{A}|-2}.
\end{equation}
Nonlinear dynamics of the nodal modes is therefore dominated by pressure coupling (we note that $|\multind{A}|\geq 3$).


\begin{table}
  \caption{Selected coupling coefficients of the epicyclic modes. 
  The second column provides formulae for the coupling coefficient
  in a general gravitational field. The third column provides the
  corresponding numerical values of the dimensionless coupling 
  coefficients for the case of the pseudo-Newtonian potential (see
  section~\ref{sec:resonances}). The radius of the torus is set to the
  location of the 3:2 epicyclic resonance, $r_0=r_{3:2}$.}             
  \label{tab:CouplingCoeffs}
  \centering          
  \begin{tabular}{l c r}   
    \hline\hline\rule{0ex}{2ex}\rule[-1.5ex]{0ex}{2ex}           
     & $\kappa$(general formula) & $\bar{\kappa}$~(PN) \\ 
    \hline\rule{0mm}{1ex}\\
    $\kappa_{\rad\,\vert\,\vert} = \kappa_{\cc{\rad}\,\vert\,\vert}^\ast$ & 
      $-\frac{1}{2}\mathcal{M}\mathcal{A}_\rad\mathcal{A}^2_\vert\varphi_{12}$ &
      $-\sqrt{2}$ \\
    $\kappa_{\rad\,\rad\,\vert\,\vert} = \kappa_{\cc{\rad}\,\cc{\rad}\,\vert\,\vert}^\ast$ &
      $\frac{1}{6}\mathcal{M}\mathcal{A}_\rad^2\mathcal{A}_\vert^2
      \left[\frac{4}{r_0\bar{\omega}_r^2}\varphi_{12} - \varphi_{22}\right]$ &
      $18.7539$ \\
    $\kappa_{\cc{\rad}\,\rad\,\vert\,\vert}$ &
      $-\frac{1}{6}\mathcal{M}\mathcal{A}_\rad^2\mathcal{A}_\vert^2  
      \left[\frac{4}{r_0\bar{\omega}_r^2}\varphi_{12} + \varphi_{22}\right]$ &
      $-11.6211$ \\
    $\kappa_{\rad\,\rad\,\rad\,\vert\,\vert}$ &
      $-\frac{1}{24}\mathcal{M}\mathcal{A}_\rad^3\mathcal{A}_\vert^2
      \left[\frac{12}{r_0^2\bar{\omega}_r^2}\left(\varphi_{12}-
      r_0\varphi_{22}\right)+\varphi_{32}\right]$ &
      $-112.754$ \\
    \rule{0mm}{1ex}\\ \hline\hline                
  \end{tabular}
\end{table}

\subsection{Epicyclic modes}

The Lagrangian eigenfunctions of the epicyclic modes are almost uniform on the torus cross-section, They cannot be coupled with each other by pressure forces in the infinitely slender torus, since they are determined by gradients in the Lagrangian displacements. However, as pointed out by \citet{Blaes+2007}, they may be coupled  by a non-slender corrections in somewhat thicker tori.

To examine this possibility we approximate the eigenfunctions of the epicyclic modes to be
\begin{eqnarray}
  \vec{\xi}_\mathrm{r,v} = \vec{\xi}_\mathrm{r,v}^{(0)} + 
  \beta\vec{\xi}_\mathrm{r,v}^{(1)} + \mathcal{O}(\beta^2),
\end{eqnarray}
where $\xi_\mathrm{r,v}^{(0)}$ are given by \citeq{eq:rad} and \citeq{eq:vert}, and both coefficients are roughly similar, $|\xi_\mathrm{r,v}^{(i)}|\sim \mathcal{A}_\mathrm{r,v}$. We calculate the leading orders for both the pressure and the gravitational coupling coefficients. The contribution of the zeroth-order eigenfunctions to the gradient in the Lagrangian displacement is highly reduced because the eigenfunctions are strictly uniform over the torus cross-section to this order of approximation. The only nontrivial corrections to the gradient are due to a small azimuthal curvature of the torus; they are however of the same order in $\beta$ as the contributions of the non-slender corrections. The pressure coupling coefficients are therefore even smaller than the gravitational coupling coefficients by a factor of
\begin{equation}
  \kappa^{(\mathrm{p})}_\multind{A}/
  \kappa^{(\mathrm{g})}_\multind{A}\sim\beta^2
\end{equation}
and the total coupling coefficients are scaled according to the gravitational contributions,
\begin{equation}
  \kappa_\multind{A}\simeq\kappa^{(\mathrm{g})}_\multind{A} \sim
  \mathcal{M}\,\Omega_0^2 r_0^2\,(\mathcal{A}/r_0)^{|\multind{A}|}
\end{equation}
We conclude that in slender tori the nonlinear coupling between the epicyclic modes is governed by gravity. 

\subsection{Axisymmetric epicyclic modes}
As demonstrated in Sect.~\ref{subsec:lin}, each epicyclic mode of the torus is accompanied by its ``complex-conjugated'' analog, whose frequency is opposite and whose eigenfunction is complex-conjugated. It is therefore useful to introduce the notation
\begin{equation}
  \omega_{\cc{A}}=-\omega_{A},
  \quad
  \vec{\xi}_{\cc{A}} = \vec{\xi}_{A}^\ast
\end{equation}
and denote complex-conjugated pairs by $(A,\bar{A})$.

The second column of Table~\ref{tab:CouplingCoeffs} contains few examples of the coupling coefficients between the axisymmetric epicyclic modes obtained by substitution of relations \citeq{eq:rad} and \citeq{eq:vert} into equations \citeq{eq:kappa3g}, \citeq{eq:kappa4g} and \citeq{eq:kappa5g}. The integrands were approximated by leading-order terms in their \mbox{$\beta$-expansions}. For a simpler notation, we introduce $\varphi_{ij} = (\partial_r^i\partial_z^j\Phi)_0$.

\section{Epicyclic resonances} 
\label{sec:resonances}

We examine nonlinear interaction between two epicyclic modes of the torus. We assume that the resulting oscillations can be described by the Lagrangian displacement of the form
\begin{equation}
  \vec{\xi}(t,\vec{x}) = 
  c_\rad(t)\,\vec{\xi}_\rad(\vec{x}) + 
  c_\cc{\rad}(t)\,\vec{\xi}_\cc{\rad}(\vec{x}) + 
  c_\vert(t)\,\vec{\xi}_\vert(\vec{x}) +
  c_\cc{\vert}(t)\,\vec{\xi}_\cc{\vert}(\vec{x}).
  \label{eq:xi-epc}
\end{equation}
Any influence of the other modes on the dynamics of these two is ignored (we return to this issue in Sect.~\ref{ssec:othermodes}). The reality of the Lagrangian displacement requires that $c_\cc{\rad}=c_\rad^\ast$ and  $c_\cc{\vert}=c_\vert^\ast$. The coefficients $c_\rad(t)$ and $c_\vert(t)$ are solutions of the two coupled-oscillator equations \citeq{eq:gov}. We assume that $c_A\sim\epsilon\ll 1$ ($\epsilon$ is a dimensionless parameter, describing the smallness of the amplitudes), then the oscillators are weakly coupled and the nonlinearities on the right-hand sides of the equations can be treated as perturbations.

Apart from the main oscillations, whose frequencies are close to the eigenfrequencies $\omega_\rad=m_\rad\Omega_0+\omega_r$ and $\omega_\vert=m_\vert\Omega_0+\omega_z$, the solutions contain harmonics that appear because of the nonlinearities in equations \citeq{eq:gov}. When the two eigenfrequencies are almost commensurable, interactions between the harmonics and main oscillation lead to resonances. The \emph{order of resonance} is given by the order of nonlinearities required to excite the resonance. For a general system with two degrees of freedom, the order of resonance $p$:$q$ ($p$ and $q$ are relative prime numbers) is given by
\begin{equation}
  n_{p:q} = p+q-1.
\end{equation}
The only exception to this simple rule is the 1:1 resonance, the order of which is $n_{1:1}=3$. 

Generally, the strength of the resonance decreases with increasing order. The resonances of higher order are more difficult to tune because the resonance range scales as $\epsilon^{n_{p:q}}\omega$ ($\omega$ is an eigenfrequency of the system). Moreover, the amplitudes and phases of resonant oscillations are modulated on the timescale proportional to $\epsilon^{n_{p:q}}\omega^{-1}$.

\subsection{Selection rules for epicyclic resonances}

In systems with some intrinsic symmetries, the presence of harmonics and occurrence of resonances depends on the symmetry properties of oscillation modes. Some resonances do not occur even though the eigenfunctions satisfy the corresponding resonance conditions because the nonlinearities involved in the production of the harmonics vanish.


\begin{table}
  \caption{Possible resonances up to the fourth order. Due to the
  equatorial-plane reflection symmetry of the equilibrium torus, 
  many resonances are absent between epicyclic modes.}             
  \label{tab:Resonances}
  \centering          
  \begin{tabular}{c c c c}   
    \hline\hline\rule{0ex}{2ex}\rule[-1.5ex]{0ex}{2ex}           
    Order &  General system &  \multicolumn{2}{c}{Epicyclic modes} \\ 
          &                 & resonance & resonance condition \\
    \hline\rule{0mm}{0.5ex}\\
    2nd & 1:2 & 1:2   & $m_\rad = 2m_\vert$\\
        & 2:1 &       &\\
    \rule{0mm}{0.5ex} &\\
    3rd & 1:3 &       &\\
        & 1:1 & 1:1   & $m_\rad = m_\vert$\\
        & 3:1 &       &\\
    \rule{0mm}{0.5ex} &\\
    4th & 1:4 & 1:4   & $m_\rad = 4m_\vert$\\
        & 2:3 &       &\\
        & 3:2 & {\bf 3:2} & $3m_\rad = 2m_\vert$\\
        & 4:1 &      \\
    \rule{0mm}{0.5ex}\\ \hline\hline                
  \end{tabular}
\end{table}

We explore possible resonances up to the fourth order using the method of multiple scales \citep{Nayfeh+Mook1979}. The result is shown in the second column of Table~\ref{tab:Resonances}. We assume that all coupling coefficients are nonzero, which corresponds to a general system with no intrinsic symmetry. Next, we apply the vertical selection rule \citeq{eq:rulez}, taking into account parities of the radial and vertical epicyclic modes, $\epsilon_{\rad,\vert}=\pm 1$. Possible resonances are listed in the third column of Table~\ref{tab:Resonances}. The fourth column shows additional conditions for azimuthal wavenumbers obtained using the azimuthal selection rule \citeq{eq:rulephi}. 

Apparently, both selection rules reduce significantly the number of possible resonances -- more of them are possible only for the axisymmetric modes ($m_\rad = m_\vert = 0$). In the strong gravitational field of both rotating and non-rotating black holes, the vertical epicyclic frequency is always greater than the radial one. Therefore the first three resonances listed in the third column of Table~\ref{tab:Resonances} does not occur for axisymmetric epicyclic modes and the strongest epicyclic resonance is therefore 3:2. We note that \citet{Kluzniak+Abramowicz2002} anticipated this result by using the analogy of the parametric resonance in the Mathieu equation. \citet{Rebusco2004} and \citet{Horak+Karas2006} achieved similar results in their discussion of internal resonances in the test-particle epicyclic motion.

As follows from the azimuthal selection rule, the necessary condition for $p$:$q$ resonance between two non-axisymmetric epicyclic modes is $p m_\rad-q m_\vert=0$ (see Table~\ref{tab:Resonances}). Therefore, the resonance condition is the same as for the axisymmetric modes, $\omega_z/\omega_r = p/q$ (and the resonance occurs therefore at the same radius), but the frequencies of the modes are different, $\omega_\vert = m_\vert\Omega\pm\omega_z$ and $\omega_\rad = m_\rad\Omega\pm\omega_r$.

\subsection{The strong-gravity 3:2 epicyclic resonance} \label{ssec:32}

We explore the strength and resonance range of the 3:2 epicyclic resonance between the radial and vertical axisymmetric epicyclic modes in a slender torus. We study parametric excitation of vertical epicyclic oscillations by radial oscillations. For simplicity, we ignore any feedback of the vertical oscillations to the radial. This is a reasonable approximation when the amplitude of the radial mode is far greater then the amplitude of the vertical mode. 

The effects of strong gravity on the central object (such as a non-rotating black hole or a compact neutron star) are included by using the \citet{Paczynski+Wiita1980} pseudo-Newtonian potential, 
\begin{equation}
  \Phi=-\frac{GM}{R-R_\mathrm{s}},
  \quad
  R=\sqrt{r^2+z^2},
\end{equation}  
where $G$ is the gravitational constant, $M$ is the mass of the compact object and $R_\mathrm{s}=2GM/c^2$ is the  Schwarzschild radius. The radial epicyclic frequency is always smaller than the vertical one. The latter is equal to the Keplerian orbital frequency because the gravitational field of the compact object is spherically symmetric. The 3:2 epicyclic resonance occurs when the torus radius is $r_0 = r_{3:2} = 9.2\,GM/c^2$ (the 3:2 radius is at $10.8\,GM/c^2$ in Schwarzschild spacetime). 

We first renormalize the coefficients $c_A$ in Eq.~\citeq{eq:gov} and introduce the dimensionless coupling coefficients $\bar{\kappa}_\multind{A}$,
\begin{equation}
  \bar{c}_A = \sqrt{\frac{\omega_A b_A}
  {\mathcal{M}\,r_0^2\Omega_0^2}} c_A,
  \quad
  \bar{\kappa}_\multind{A} = \left(\prod_{k=1}^{|\multind{A}|}
  \sqrt{\frac{\mathcal{M}\,r_0^2\Omega_0^2}{\omega_{A_k} b_{A_k}}}
  \right)\frac{\kappa_\multind{A}}{\mathcal{M}\,r_0^2\Omega_0^2}.
\end{equation}
The governing equations then take the form
\begin{equation}
  \der{\bar{c}_A}{t} + \ii\omega_A \bar{c}_A=  
  \ii\omega_A \bar{\mathcal{F}}^\ast_A(\bar{c}_I),
  \label{eq:gov-norm}
\end{equation}
where the formula for the dimensionless nonlinear functions $\bar{\mathcal{F}}_A(\bar{c}_I)$ is analogous to Eq.~\citeq{eq:nonlin} (all quantities are indicated by a bar). Numerical values of several dimensionless coupling coefficients evaluated at $r_0=r_{3:2}$ are listed in the third column of Table~\ref{tab:CouplingCoeffs}.


\begin{figure}[t]
  \begin{center}
    \includegraphics[width=0.48\textwidth]{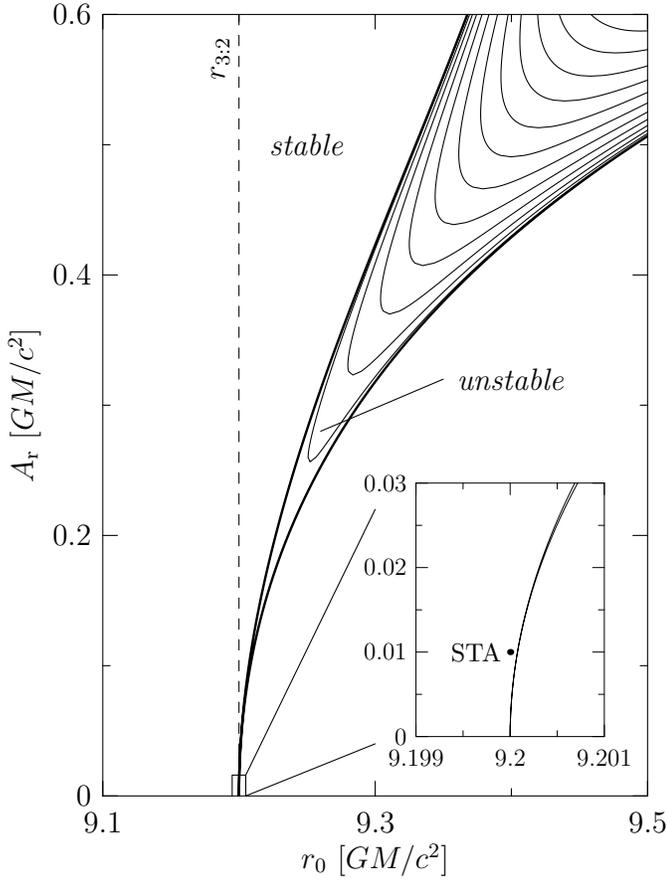}
  \end{center}
  \caption{Region of the 3:2 parametric-like resonance between the epicyclic modes in pseudo-Newtonian slender tori. The resonance region is shown in the plane of the torus position (horizontal) versus the radial epicyclic mode amplitude (vertical). The domains where the vertical epicyclic oscillations are stable and unstable are separated by transition curves. The contours denote constant growth-rate of vertical oscillations. The corresponding values of $\gamma/\Omega_0$ are multiples of $0.25\times 10^{-2}$ (zero corresponds to the transition curves). The inset of the figure shows the position and radial amplitude of the torus (``STA''-point) in the simulations of \citet{Sramkova+2007}.}
  \label{fig:tongue}
\end{figure}

The solution of equation \citeq{eq:gov-norm} can be found using the method of multiple scales. The radial oscillations can be approximated by $\bar{c}_\rad=\bar{A}_\rad\exp(-\ii\omega_r t)$. Without any loss of generality, we assume that the initial phase of the oscillations is such that $\bar{A}_\rad\in\mathbb{R}$. Up to the leading order, the vertical oscillations are given by $\bar{c}_\vert=\bar{A}_\vert(t)\exp(-\ii\omega_z t)$, where $\bar{A}_\vert(t)\in\mathbb{C}$ is a slowly changing amplitude. The slow time evolution is given by the amplitude equation, the form of which is 
\begin{equation}
  -\frac{\ii}{\omega_z}\der{\bar{A}_\vert}{t} + 
  \left(\lambda_\rad |\bar{A}_\rad|^2 + 
  \lambda_\vert |\bar{A}_\vert|^2\right) \bar{A}_\vert +
  \alpha\,\bar{A}_\vert^\ast \bar{A}_\rad^3 \ee^{\ii\delta\omega t} = 0,
  \label{eq:amp-complex}
\end{equation}
where $\delta\omega = 2\omega_z - 3\omega_r$ describes the detuning of the eigenfrequencies from the rational ratio. It depends on the location of the torus center $r_0$ ($\delta\omega=0$ when $r_0=r_{3:2}$). The dimensionless constants $\alpha$, $\lambda_\rad$ and $\lambda_\vert$ are given by coupling coefficients. The corresponding formulae are
\begin{eqnarray}
  \alpha &=& \frac{1}{2}\omega_z\Big[
  72\bar{\kappa}_{\cc{\rad}\vert\vert}^2\bar{\kappa}_{\rad\vert\vert} + 
  9\bar{\kappa}_{\cc{\rad}\vert\vert}\left(
  \bar{\kappa}_{\rad\vert\vert}^2 - 4\bar{\kappa}_{\rad\rad\vert\vert}\right) - 
  \nonumber\\ &\phantom{=}&
  9\bar{\kappa}_{\rad\vert\vert} 
  (\bar{\kappa}_{\cc{\rad}\cc{\rad}\vert\vert} + \bar{\kappa}_{\rad\rad\vert\vert}) -
  8\bar{\kappa}_{\rad\rad\rad\vert\vert}
  \Big],
\end{eqnarray}
\begin{equation}
  \lambda_\rad = \frac{3}{2}\omega_z\left[
  \bar{\kappa}_{\cc{\rad}\vert\vert}^2 + 2\bar{\kappa}_{\rad\vert\vert}^2 - 
  4\bar{\kappa}_{\cc{\rad}\vert\vert}
  \right],
\end{equation}
\begin{equation}
  \lambda_\vert = 3\omega_z\bar{\kappa}_{\rad\rad\vert\vert}.
\end{equation}
Numerical values of these constants at the resonance radius are $\alpha=1101$, $\lambda_\rad=78.14$ and $\lambda_\vert = 1.333$.

When the amplitude of the vertical oscillations is small, the second term in brackets in Eq.~\citeq{eq:amp-complex} can be neglected with respect to the first one and we obtain a linear equation. We attempt to find its solution in the form
\begin{equation}
  \bar{A}_\vert=[a_1(t)+\ii a_2(t)]\,\ee^{-\ii\delta\omega\,t/2}
\end{equation}
with two unknown real functions $a_1(t)$ and $a_2(t)$. Substituting these expressions into equation \citeq{eq:amp-complex}, we obtain 
\begin{eqnarray}
  \der{a_{1}}{t} + \left[\delta\omega + \omega_z
  \left(\lambda_\rad \bar{A}_\rad^2 + \alpha A_r^3\right)\right]a_{2}&=&0,
  \label{eq:amp-real1} \\
  \der{a_{2}}{t} + \left[\delta\omega + \omega_z
  \left(\lambda_\rad \bar{A}_\rad^2 - \alpha A_r^3\right)\right]a_{1}&=&0.
  \label{eq:amp-real2}
\end{eqnarray}
These linear equations have constant coefficients, therefore, we may assume that $a_{1,2}$ depends on $t$ as $a_{1,2}=\tilde{a}_{1,2}\exp(\gamma t)$. When this form is inserted into equations \citeq{eq:amp-real1} and \citeq{eq:amp-real2} we obtain a couple of linear homogenous algebraic equations for $\tilde{a}_{1,2}$ that are solvable when
\begin{equation}
  \gamma=\pm\left[\alpha^2\omega_z^2 \bar{A}_\rad^6 - 
  \left(\frac{1}{2}\delta\omega + 
  \omega_v\lambda_\rad \bar{A}_\rad^2\right)^2\right]^{1/2}
  \label{eq:growth}
\end{equation}
The amplitude of the vertical oscillation grows exponentially when the frequency detuning is in the range $\omega_1<\delta\omega<\omega_2$ with
\begin{equation}
  \omega_{1,2} = -2\omega_z\left(\lambda_\rad \bar{A}_\rad^2 \pm 
  \alpha \bar{A}_\rad^3\right).
  \label{eq:TransitionCurves}
\end{equation}
The region of resonance is shown in Fig.~\ref{fig:tongue}. The transition curves separating the stable and unstable regions are given by Eq.~\citeq{eq:TransitionCurves}. For a given amplitude of the radial oscillations, the size of the resonance range $\Delta\omega = |\omega_2 - \omega_1|$ is proportional to $\bar{A}_\rad^3$. Instead of the frequency detuning $\delta\omega$, we use the corresponding radial coordinate $r_0$ of the torus center in the figure ($r_0$ is calculated from the well-known dependences of the epicyclic frequencies on the radius). 

The growth-rate of the unstable vertical mode is given by Eq.~\citeq{eq:growth} and indicated by contours in Fig.~\ref{fig:tongue}. The maximal growth-rate that can be obtained for a given amplitude of radial oscillations is proportional to $\bar{A}_\rad^3$ and occurs when $\delta\omega=-2\omega_z\lambda_\rad \bar{A}_\rad^2$. It is given by $\gamma_\mathrm{max}=\alpha\omega_z \bar{A}_\rad^3$. In the figure, the amplitude of the radial epicyclic oscillations is shown in the units of $GM/c^2$. The value of $A_\rad$ is connected to the value $\bar{A}_\rad$ of the dimensionless amplitude by
\begin{equation}
  A_\rad = 2\xi_\rad^r\sqrt{\frac{\mathcal{M} r_0^2\Omega^2}{\omega_\rad b_\rad}}\bar{A}_\rad = 
  \frac{\sqrt{2}r_0}{\bar{\omega}_r} \bar{A}_\rad,
\end{equation}
where $\xi_\rad^r$ is the radial component of the radial epicyclic-mode eigenfunction [Eq.~\citeq{eq:rad}].

\section{Discussion} 
\label{sec:discusion}
\subsection{Strength of the resonance}

We have found that the 3:2 epicyclic resonance is very sensitive to the precise tuning of the eigenfrequencies of the torus. For small amplitudes of the radial oscillations, Eq.~\citeq{eq:TransitionCurves} implies that  the range of the detuning parameter, for which the resonance may operate, is very limited. Moreover, the growth-rate of the vertical epicyclic oscillations in the resonance is quite small. It appears that these facts make the epicyclic  resonance difficult to observe in both, numerical simulations as well as real astrophysical objects.

In principle, our results agree with the numerical simulations of \citet{Sramkova+2007}, who claim that the epicyclic modes are not resonantly coupled. In their simulations the radial extent of the torus  is smaller than its radius by the factor $\sim 0.02$, while the velocity amplitude of the radial perturbations are smaller than the central sound speed by the factor $\sim 0.3$, This situation corresponds to the amplitude of the radial oscillations $A_\rad\sim10^{-2} GM/c^2$. In the simulations, the torus center is at $r_0=9.2 GM/c^2$ (models A3 and A4). The corresponding point $[r_0,A_\rad]$ is outside the resonance tongue as shown in Fig.~\ref{fig:tongue}. For the same amplitude of radial oscillations our theory predicts maximal growth-rate $\gamma_\mathrm{max}\sim 10^{-7}\Omega_0$. Besides the high precision of the calculations, an observation of the resonance in numerical simulations requires considerably long simulation times and a precisely tuned radius of the torus.

\subsection{Comparison of the characteristic time-scales}

In contrast to the inviscid flow considered in this paper, real accretion tori are made of viscous fluids. Low viscosity causes slow secular evolution of the torus by changing its angular-momentum distribution on a viscous timescale. These changes affect the torus radius $r_0$. Similarly, a viscous heating causes secular evolution of the torus volume on the thermal timescale. Both processes influence the eigenfrequencies of the torus and may therefore disturb the precise tuning, which is required by the resonance. This happens when the resonant modulation timescale $1/\gamma$ exceeds the characteristic time that the torus spends in the resonance region. We adopt common formulae for a thin disk \citep{Frank+1992}
\begin{eqnarray}
  t_\mathrm{visc}&=&\alpha_\mathrm{t}^{-1}(H/R)^{-2}\,t_\mathrm{orb} 
  \sim \alpha_\mathrm{t}^{-1}\beta^{-2}\,t_\mathrm{orb} , 
  \\
  t_\mathrm{th}&=&\alpha_\mathrm{t}^{-1} t_\mathrm{orb},
\end{eqnarray}
where $\alpha_\mathrm{t}$ parameterize the accretion disk turbulence \citep{Shakura+Syunyaev1973} and $t_\mathrm{orb}\sim\Omega_0^{-1}$ is the orbital timescale. The typical drifts of eigenfrequencies on these timescales due to the changing radius or the volume of the torus are $(\Delta\omega)_\mathrm{visc}\sim\Omega_0$ and  $(\Delta\omega)_\mathrm{th}\sim\beta^2\Omega_0$. Therefore, typical times that the torus spends in the resonance region undergoing viscous or thermal changes are similar, $(\Delta T)_\mathrm{visc}\sim(\Delta T)_\mathrm{th}\sim \alpha_\mathrm{t}^{-1}\,\beta^{-2}\alpha(A_\rad/r_0)^3\,t_\mathrm{orb}$. If we compare this result with the resonant  timescale, we obtain an upper limit on $\alpha_\mathrm{t}$ for the resonance to operate,
\begin{equation}
  \beta^2\alpha_\mathrm{t}\lesssim\alpha^2\left(\frac{A_\rad}{r_0}\right)^{6}.
  \label{eq:alpha}
\end{equation}

\subsection{Effects of other oscillation modes} \label{ssec:othermodes}

We considered nonlinear interactions between two epicyclic modes and ignored the influences of all other modes [see Eq.~\citeq{eq:xi-epc}]. If they are present with small amplitudes in the oscillations, the Eq.~\citeq{eq:amp-complex} is modified by the presence of additional terms $\lambda_I|A_I|^2$ in brackets. The shape of the transition curves and the growth-rates of the vertical epicyclic modes are then slightly changed, however the size of the resonance range and the general discussion presented in this paper remain the same.

More importantly, the epicyclic resonance may be suppressed by the parametric instability that may quickly advect the energy from the epicyclic modes to some low-frequency modes of the torus \citep{Dziembowski1982, Wu+Goldreich2001, Arras+2003, Nowakowski2005}. Being a resonant interaction among three modes, the characteristic timescale of the parametric instability is far shorter than that of the 3:2 epicyclic resonance. An important condition for this process is the existence of a pair of low-frequency modes that form a resonant triple with the epicyclic modes. This happens when the frequencies and azimuthal wavenumbers of the low-frequency modes satisfy  the conditions $\omega_{r,z} \approx \omega_1 + \omega_2$ and $m_1 + m_2=0$. The `parent' mode (here radial or vertical axisymmetric epicyclic mode) of the highest frequency $\omega_{r,z}$ is a source of energy for the two `daughter' modes with frequencies $\omega_{1,2}$. For each pair of daughter modes, there exists a lower limit to the parent-mode amplitude above which the parametric instability begins to operate. This limit depends on the damping rates $\gamma_1$, $\gamma_2$ of daughter modes and on a coupling coefficient $\kappa$ of the three-mode interaction as $A_\mathrm{min}\propto\gamma_1\gamma_2/|\kappa|^2$ \citep{Dziembowski1982}.

To decide whether the parametric instability advects energy from the epicyclic modes to some other low-frequency (perhaps unobserved) modes, it is necessary to explore in details the eigenfrequencies, eigenfunctions and damping rates of the torus modes. Such analyses has been carried out by \citet{Wu+Goldreich2001} and \citet{Arras+2003} in the context of stellar pulsations. A similar study is beyond the scope of this paper because the damping processes in tori are still only purely understood. We therefore summarize the necessary properties of potential daughter modes, based on known properties of the eigenfrequencies and eigenfunctions of the slender torus.

We first examine whether possible pairs of daughter modes exist among the lowest order modes of the constant angular-momentum tori derived by \citet{Blaes+2006} (see their Table 4). Both axisymmetric and non-axisymmetric modes with $m=1$ and $2$ are considered (the modes with higher azimuthal wavenumber are excluded because their  frequencies are higher than $\omega_{r,z}$). We evaluate the eigenfrequencies in the pseudo-Newtonian gravitational field at $r=r_{3:2}$ and select only those pairs whose mutual interactions with the epicyclic modes are not forbidden by the selection rules. For each such combination, we evaluate the frequency detunings $\Delta\omega_{r,z} = \omega_{r,z} - \omega_1 - \omega_2$. We do not find any pair that provides $\Delta\omega_{r,z}$ within 10\% of the orbital velocity $\Omega_0$. The situation is probably similar for steeper angular-momentum distributions. We expect that the resonance condition will be satisfied only for some particular values of $\hat{\kappa}$. Therefore, we may conclude that the resonant coupling is inefficient and the parametric instability does not drive the energy from the epicyclic to the lowest-order modes of the torus. 

The high-order modes, whose eigenfunctions have many nodes in both the $r$ and $z$ directions, can be treated using the WKBJ approximation. These modes are damped more strongly and therefore they advect energy from the epicyclic modes more effectively. Their eigenfunctions can be approximated by
\begin{equation}
  \vec{\xi} = \hat{\vec{\xi}}\,\ee^{\ii(m\phi + \vartheta)},
  \quad
  \vartheta = \int k_r\dd r + \int k_z\dd z,
\end{equation}
where $k_r, k_z \gg r_0^{-1}$ and $\hat{\vec{\xi}}$ are slowly varying functions of $r$ and $z$ and $\vartheta$ is a rapidly changing phase. The wave-vector $(k_r,k_z)$ is connected to the eigenfrequency of the mode by the dispersion relation that takes the form (i) $\sigma^2 = c_\mathrm{s}^2(k_r^2 + k_z^2) $ for the acoustic and surface-gravity modes ($c_\mathrm{s}$ is local speed of sound) and (ii) $\sigma^2 = [k_z^2/(k_r^2 + k_z^2)]\hat{\kappa}^2$ for the internal-inertial modes \citep{Blaes+2006}. 

We suppose that the two high-order daughter modes, 1 and 2, are governed by the same dispersion relation. The integrands of the three-mode coupling coefficients $\kappa_{\rad12}$ and $\kappa_{\vert12}$ contain an exponential $\exp[\ii(\vartheta_1 + \vartheta_2)]$ (the epicyclic modes are almost uniform, therefore their contributions are negligible). The integrand contributes significantly only close to the point where the phase of the exponential becomes stationary in both $r$ and $z$ directions and therefore $k_{1r}+k_{2r} \approx k_{1z}+k_{2z}\approx0$. Using the dispersion relation, we conclude that this happens when $\sigma_{1} \approx \pm\sigma_{2}$. On the other hand, the two modes form a resonant triple with an epicyclic mode when $\omega_{r,z}\approx\sigma_1+\sigma_2$. Hence, the corotation frequencies of the potential daughter modes are  $\sigma_1\approx\sigma_2\approx\omega_{r,z}/2$, which corresponds to the eigenfrequencies $\omega_{1,2}\approx\omega_{r,z}/2\pm m\Omega_0$. The parametric instability operates only when the mode offering the energy is the one with the highest frequency, i.e.\ when $\omega_{1,2}<\omega_{r,z}$. The only possible value of the azimuthal wavenumber is then $m=0$.

The frequencies of the acoustic and surface-gravity modes that are governed by the dispersion relation (i), increase with increasing $k_r$ and $k_z$. Therefore these modes will not drain energy from the epicyclic modes. Consequently, the three-mode parametric instability is not dangerous for the epicyclic resonance in constant angular-momentum tori. 

The frequencies of the inertial oscillations governed by relation (ii) are, however, always between 0 and $\hat{\kappa}$. As the angular-momentum distribution approaches the Keplerian one, $\hat{\kappa}$ approaches the radial epicyclic frequency. Therefore, for sufficiently steep distribution of angular momentum, there will be high-order axisymmetric inertial modes whose eigenfrequencies are sufficiently close to $\omega_{r,z}/2$. These modes may be important for the parametric instability. Further careful analysis is required in order to determine whether the nonlinear interactions with these modes is sufficient to suppress the epicyclic resonance.

\section{Conclusions} 
\label{sec:conclusions}
We have applied a general theory of nonlinear pulsation in rotating stars to a problem of nonlinear oscillations of thick accretion disks. In this note, we have calculated the strength of the coupling between epicyclic modes of slender torus. We have taken a closer look to a parametric excitation of vertical epicyclic motion of the torus due to radial epicyclic oscillations in the epicyclic resonance. Our main findings can be summarized as follows.

(1)
While the nodal modes of the torus are coupled by pressure gradients, the coupling between the epicyclic modes is governed by gravity (see Sect.~\ref{sec:cplcoeff}).

(2)
The strongest epicyclic resonance between the axisymmetric modes occurs when the ratio of the vertical and horizontal epicyclic frequencies are close to the 3:2 ratio. In addition, the 3:2 resonance may appear also between two non-axisymmetric modes, whose azimuthal wavenumbers satisfy $3m_\rad = 2m_\vert$. However, already the first possible resonance of this kind ($m_\rad = 2$ and $m_\vert = 3$) provides frequencies that are apparently too high to be identified with those observed in QPOs (Sect.~\ref{sec:resonances}).

(3)
The azimuthal selection rule forbids any resonance between one axisymmetric and one non-axisymmetric epicyclic mode. Note that such resonance has been recently suggested by \citet{Bursa2005} to explain discrepancy between spin estimates based on spectral fitting and resonance models. We note that the conclusions (3) and (4) are consequences of the axial and equatorial-plane symmetry of the equilibrium torus only (Sect.~\ref{sec:resonances}).

(4) We have examined the stability of the vertical epicyclic motion in the 3:2 epicyclic resonance with the radial epicyclic oscillations. For a given amplitude of the radial oscillations we have found a range of torus radii for which the vertical oscillations are unstable. We have noted that the epicyclic resonance become significant only for high amplitudes of the radial oscillations (see Sect.~\ref{ssec:32}).


\begin{acknowledgements}
I appreciate fruitful discussions with Paola Rebusco, Eva \v{S}r\'{a}mkov\'{a}, Marek Abramowicz, Wlodek Klu\'{z}niak, Omer Blaes, Gabo T\"{o}r\"{o}k and colleagues from the Astronomical institute in Prague. I am also grateful to an anonymous referee for many valuable comments that largely improved the manuscript. Finally, I  acknowledge the hospitality of MPI Garching and financial support of the GA\v{C}R grant 205/06/P415 and of the Center for Theoretical Astrophysics (LC\,06014). 
\end{acknowledgements}


\bibliographystyle{aa}
\bibliography{epicpl}


\end{document}